# A Binary Classifier-Based Wire Resistance Attack on the KLJN Secure Key Exchanger

MEHMET YILDIRIM[§], FAHRETTIN AY, LASZLO B. KISH

[1]*Department of Electrical and Computer Engineering, Texas A&M University,*
*College Station, TX 77841-3128, USA*
*eem.mehmetyildirim@tamu.edu, laszlokish@tamu.edu*

**Abstract:** The statistical fluctuations of the mean-square noise voltages measured at Alice's and Bob's ends in the KLJN scheme are used to implement a binary classifier for a new type of wire-resistance-based attack. The data are plotted on a two-dimensional graph, where the x- and y-axes represent the mean-square voltages at Alice's and Bob's ends, respectively. When the wire resistance is nonzero, the data form distinct lines for the LH and HL cases, allowing Eve to extract the secure bits with nearly 100% success. Further analysis shows that swapping the x and y axes for the LH data reproduces the curve for the HL case, effectively reducing the number of independent measurements by half. These results suggest that machine learning tools could exploit this property for enhanced detection performance, although such methods are unnecessary here since the LH and HL cases are completely separable. The only effective defense against this attack remains the traditional approach: properly increase the noise temperature on the side with lower resistance, or equivalently, scale down the noise temperature on the higher-resistance side. All claims are confirmed through computer simulations.

## 1. Introduction

Information-theoretic security, that is, unconditional (information-theoretic) security, is a notion of security where a cryptosystem remains secure even against an adversary with unlimited computational power, because the ciphertext reveals no information about the plaintext [1-4]. The foundation component of a system with unconditional security is the secure key exchange. The security of a cryptographic communication scheme cannot be better than the security of its key exchange protocol.

So far, only two hardware schemes (and no software schemes) exist that can claim unconditional security: quantum key distribution (QKD) [5-43] and the Kirchhoff-Law-Johnson-Noise (KLJN) protocol, which is a classical physical counterpart of QKD [44-103].

The KLJN secure key exchanger is a classical statistical physical key distribution scheme proposed in 2005 [3,44,45]. It offers information-theoretic (unconditional) security based on classical thermal noise and Kirchhoff's circuit laws rather than quantum mechanics. Its security proof is given (among others) in [46]. The KLJN system uses pairs of resistors representing binary values (commonly labeled as $R_L$ and $R_H$) and Gaussian Johnson noise voltage sources to mask the voltage and current signals transmitted through a shared wire. During each clock period, Alice and Bob randomly select resistors, and the cases where they choose different values, $R_L/R_H$ (that is LH) or $R_H/R_L$ (that is HL), to form secure key values since an eavesdropper cannot distinguish these two configurations solely through passive observation, as they have identical noise characteristics, see the explanation as follows.

[§] Corresponding Author

The fundamental mathematical expression showing why the mixed resistor states $R_A \neq R_B$ in the KLJN protocol are indistinguishable to an eavesdropper involves the mean-square voltage on the wire. For two resistors $R_A$ and $R_B$ connected in parallel, driven by noise voltages at temperature $T$, the mean-square voltage over the wire line is given by the Johnson noise formula combined:

$$\left\langle U_w^2(t) \right\rangle = 4kT\Delta f \frac{R_L R_H}{R_L + R_H} \tag{1}$$

Here, $k$ is the Boltzmann's constant, $T$ is the absolute temperature, and $\Delta f$ is the noise bandwidth. Crucially, this expression is *symmetric* in $R_A$ and $R_B$, meaning

$$\left\langle U_w^2(t) \right\rangle_{R_L/R_H} \equiv \left\langle U_w^2(t) \right\rangle_{LH} = \left\langle U_w^2(t) \right\rangle_{HL} \equiv \left\langle U_w^2(t) \right\rangle_{R_H/R_L} \tag{2}$$

Therefore, an eavesdropper measuring the voltage noise on the line cannot distinguish between the mixed resistor configurations LH or HL because both produce exactly the same noise voltage distribution on the line. This symmetry underpins the security of the KLJN key exchange during mixed resistor states.

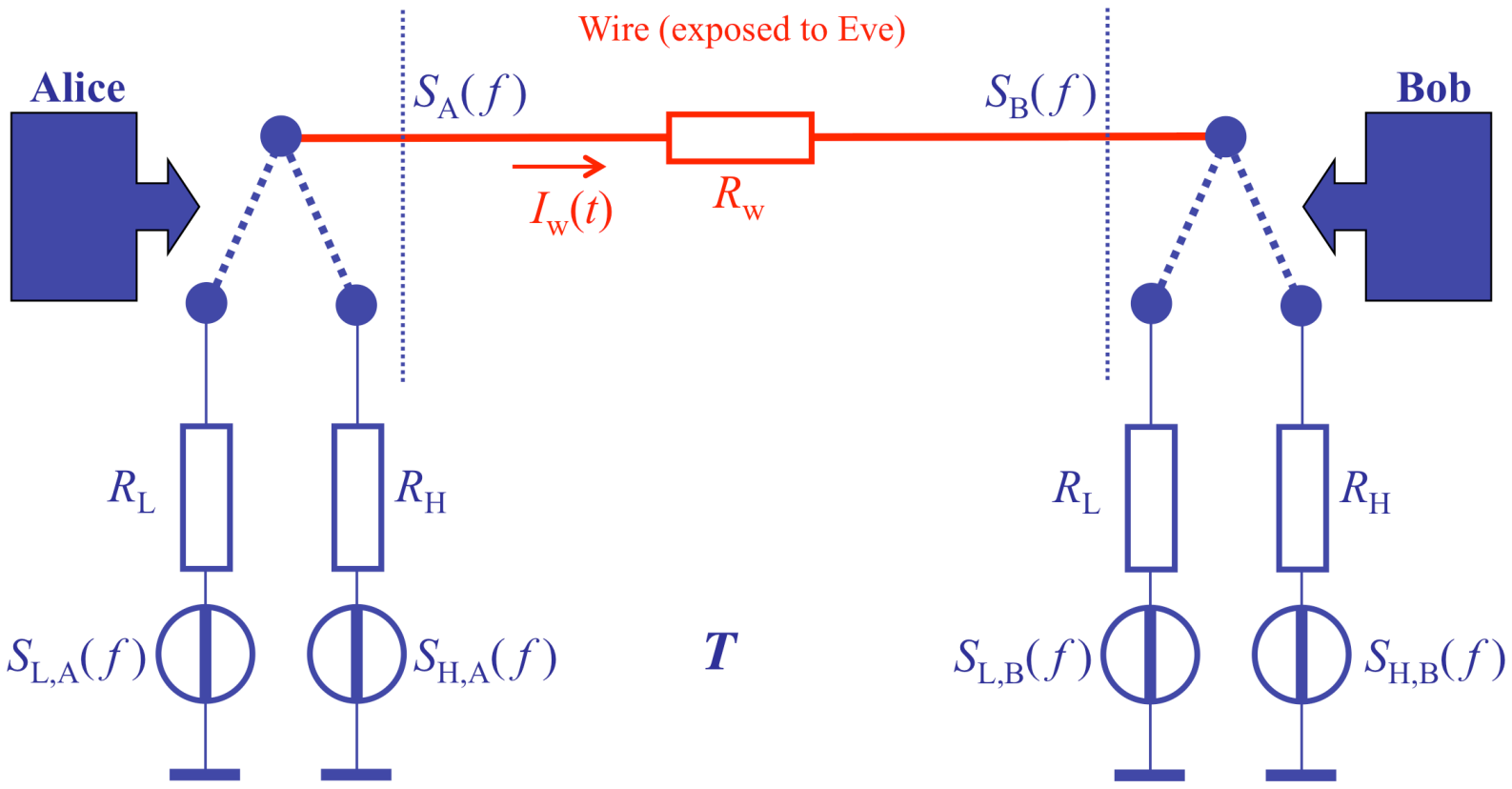

Figure 1. The KLJN scheme with non-zero wire line resistance $R_w$. $T$ is the common (effective) temperature, the four independent noise voltage generators are the thermal noises of the resistors (they are typically emulated by active generators for enhanced temperature); $S_{L,A}$=$S_{L,B}$ and $S_{H,A}$=$S_{H,B}$ are their noise spectra, and $I_w$ is the current in the wire line.

However, the wire resistance attack, one of the earliest and most discussed vulnerabilities, exploits the presence of non-zero wire resistance between the two communication parties (earliest by Janos Bergou reported by Cho in [45]). In an ideal KLJN system, the connecting wire is assumed to be perfectly resistance-free so that voltage and current fluctuations at both ends are indistinguishable. In practice, the finite wire resistance $R_w$ slightly alters the mean-square voltage at each end, lower at the end with $R_L$ and higher at the end with $R_H$.

This attack, sometimes called the Bergou–Scheuer–Yariv attack [45,47], was (due to a calculation error) initially thought to produce significant information leakage but that was later shown experimentally by Mingesz, et al, [49], and theoretically by Kish and Scheuer [47] that the actual leak is well controllable even without using privacy amplification [62].

Further work demonstrated that wire resistance effects could be fully compensated by properly boosting the noise temperature or voltage amplitude of the noise generator at the $R_L$ end, or alternatively, decreasing it at the $R_H$ end. The method was first proposed in 2014 by Kish and Granqvist (KG) [89] however their result had a calculation error and, in 2016, Vadai, Gingl, Mingesz (VGM) [52], showed the corrected formula for the temperature scaling with further applications.

According to the corrected VGM formula, the temperature $T_L = \beta T$ at the $R_L$ side is scaled by the factor:

$$\beta = \frac{1 + \frac{R_C}{2R_\mathrm{L}}}{1 + \frac{R_C}{2R_\mathrm{H}}} \quad . \tag{3}$$

In the original KG version [89], the factor 2 in the numerator and denominator were absent. The same principle was also proposed as a defense against the Second Law Attack [89], which exploits the information leak caused by power flow into the wire resistance. VGM confirmed the corrected formula through computer simulations [52].

For the application of the VGM defense principle against our new attack below, see section 2.3.

## 2. The new classifier-based attack

Traditional analysis assumes that any eavesdropper (Eve) compares the mean-square noise voltages or currents measured during the bit exchange period (BEP) at the two ends of the wire to infer which resistance configuration (LH or HL) is active. However, a more detailed examination of the fluctuations in the mean-square values provides a deeper attack mechanism.

### *2.1 The new attack protocol*

The mean-square noise voltages at Alice's and Bob's ends are recorded as $\langle V_A^2(t)\rangle = x$ and $\langle V_B^2(t)\rangle = y$ over a large number of BEPs. The set of pairs $\{x_i\ ,\ y_i\ \}$ can be plotted as a two-dimensional point pattern, with $x$ on the X-axis and $y$ on the Y-axis. Then the same procedure can be applied by repeating the process so that Alice and Bob are swapped.

In the ideal KLJN system, where the wire resistance is zero, the distributions of these points for the LH and HL states form a straight line with a slope of 1, see simulation result in Figure 2.

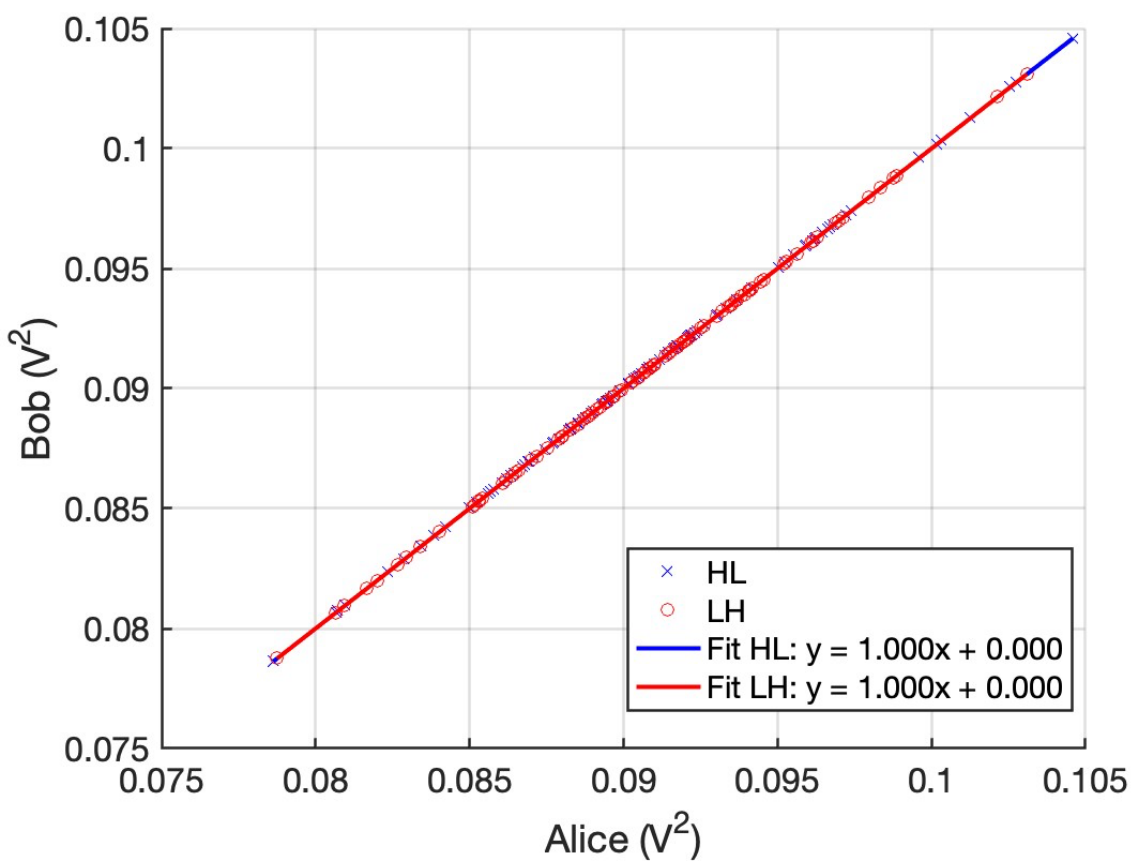

Figure 2. Statistical fluctuations at Alice's and Bob's sides in the binary qualifier plot at zero wire resistance. In the absence of voltage division by the wire resistance, the curves have the same slope, and the unity cross-correlation between Alice's and Bob's ends keeps the data points tightly on the line.

In contrast, finite wire resistance introduces two distinct fitted lines deviating from the slope of 1 due to the voltage divider formed by the wire resistance and the resistances of Alice and Bob (see Section 2.2). Additionally, the wire resistance separates Alice's and Bob's sides, which allows the data to scatter because the noises in Alice's and Bob's generators are independent. When the scattering of the points in the LH and HL cases is smaller than the separation between the fitted lines so that the two point sets do not overlap, this phenomenon can serve as a binary classifier between the LH and HL situations.

Due to Kerckhoffs's principle, Eve can simulate the given KLJN system and construct such a classifier plot. The attack protocol is then simple: Eve measures the mean-square voltage during the BEP of a bit and adds the resulting single point to the classifier plot. This new point will appear among the points of either the LH or the HL set, respectively.

Eve's preparation of the protocol can, of course, be simplified to the original recording: plotting the 2D data once, then swapping the x and y axes and replotting the data to get the second fitted line.

This two-dimensional statistical approach provides Eve with a larger information leak than the conventional one-dimensional mean-square voltage comparison, which relies solely on a single mean-square voltage value.

### *2.2 Computer simulations of the attack*

We have simulated the above protocol with Gaussian white noise, with the following parameters: sampling frequency 20 kHz; bandwidth $\Delta f = 10$ kHz; effective temperature $T$=2*10$^{14}$ K; $R_L$=1000 Ω , $R_H$=9000 Ω at various cable resistances, and BEP=100 (practical duration value, defined by the number of independent data in it).

In Figure 3, the binary classifier plot is shown for a practical case with a 200 Ω cable resistance (2% of the total wire resistance, see [49]. The coordinates of the points are generated by the mean-square voltage values at Alice's and Bob's sides. The lines are fitted by Matlab's curve fitting algorithm. The data points closely follow the lines without overlap between the LH and HL cases, indicating a practically 100% success rate when Eve uses this classifier for cracking the secure bits.

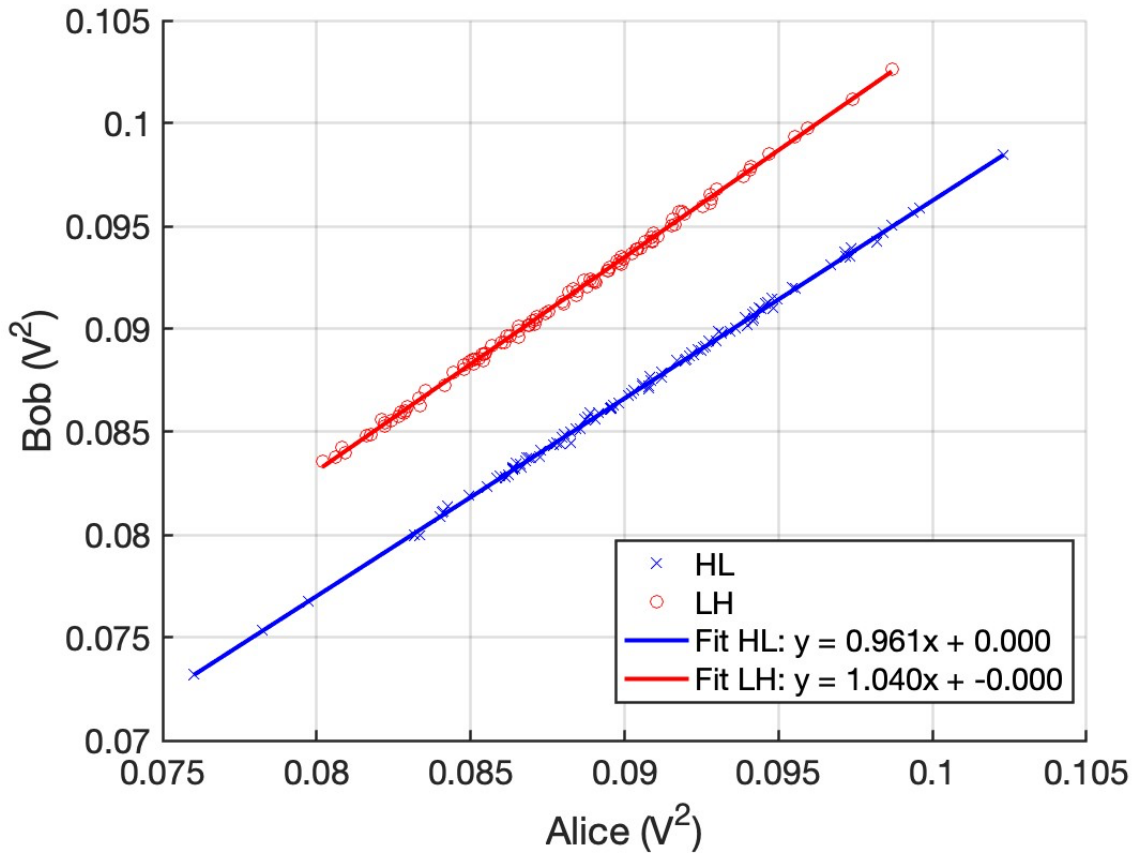

Figure 3. Binary classifier plot for a practical case with a 200 Ω cable resistance [2% of the total wire resistance; see Ref. [49] and a practical BEP = 100 value. The data points closely follow the lines without overlap between the LH and HL cases, indicating a practically 100% success rate in cracking the secure bits.

In Figure 4, the binary classifier plot is shown for a hypothetical case with a large cable resistance of 800 Ω. The data points scatter away from the lines much more than earlier because the cross-correlation between Alice’s and Bob’s sides weakens as the wire resistance increases. Nevertheless, due to the greater separation between the lines, there is still no overlap between the LH and HL cases, indicating again a practically 100% success rate when the classifier plot is used for cracking the secure bits.

Finally, Figure 5 shows how well swapping the x and y axes for the LH data reproduces the curve for the HL case, thus reducing the number of independent simulations needed to construct the classifier; compare the slopes with the values in Figure 3. Once the classifier has been prepared, Eve needs only a single measurement per bit and to plot the corresponding point on this graph to determine the secure bit value.

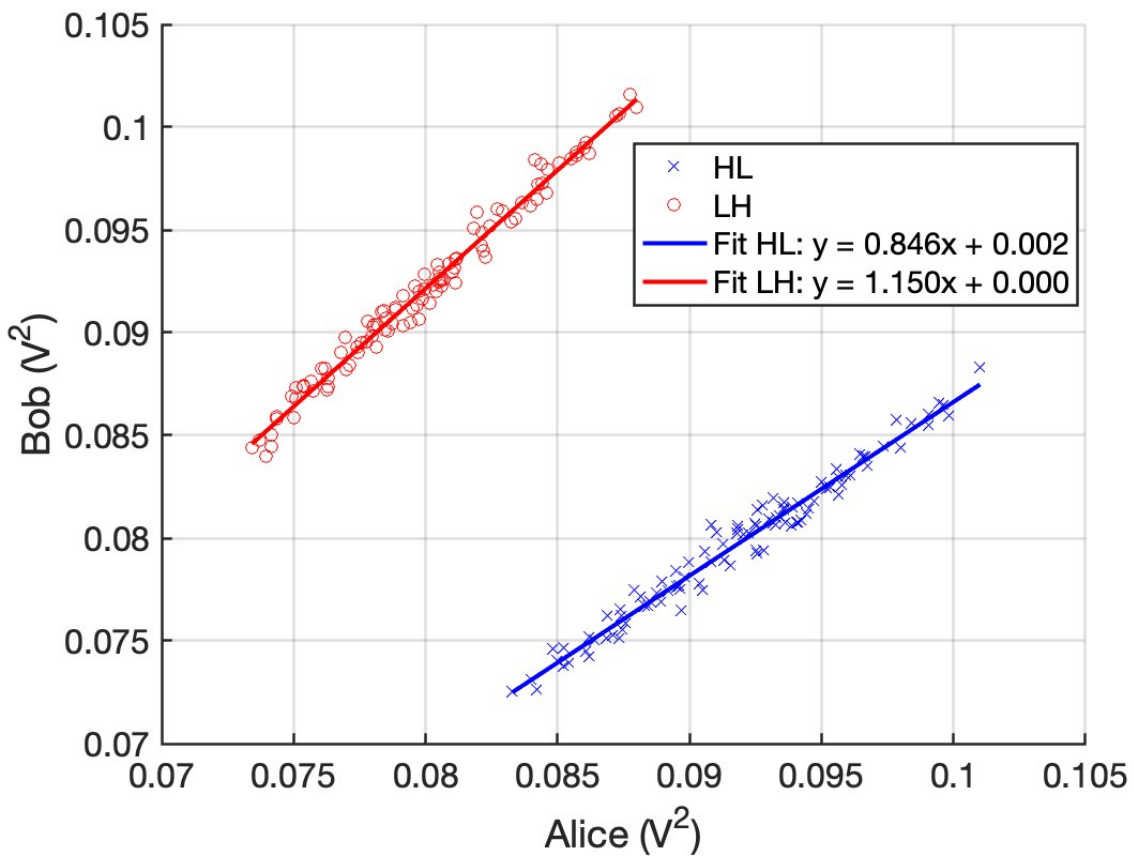

Figure 4. Binary classifier plot for a hypothetical case with a large cable resistance of 800 Ω and a practical BEP = 100 value. The data points scatter away from the lines because the cross-correlation between Alice's and Bob's sides weakens as the wire resistance increases. Nevertheless, due to the greater separation between the lines, there is still no overlap between the LH and HL cases, indicating a practically 100% success rate in cracking the secure bits.

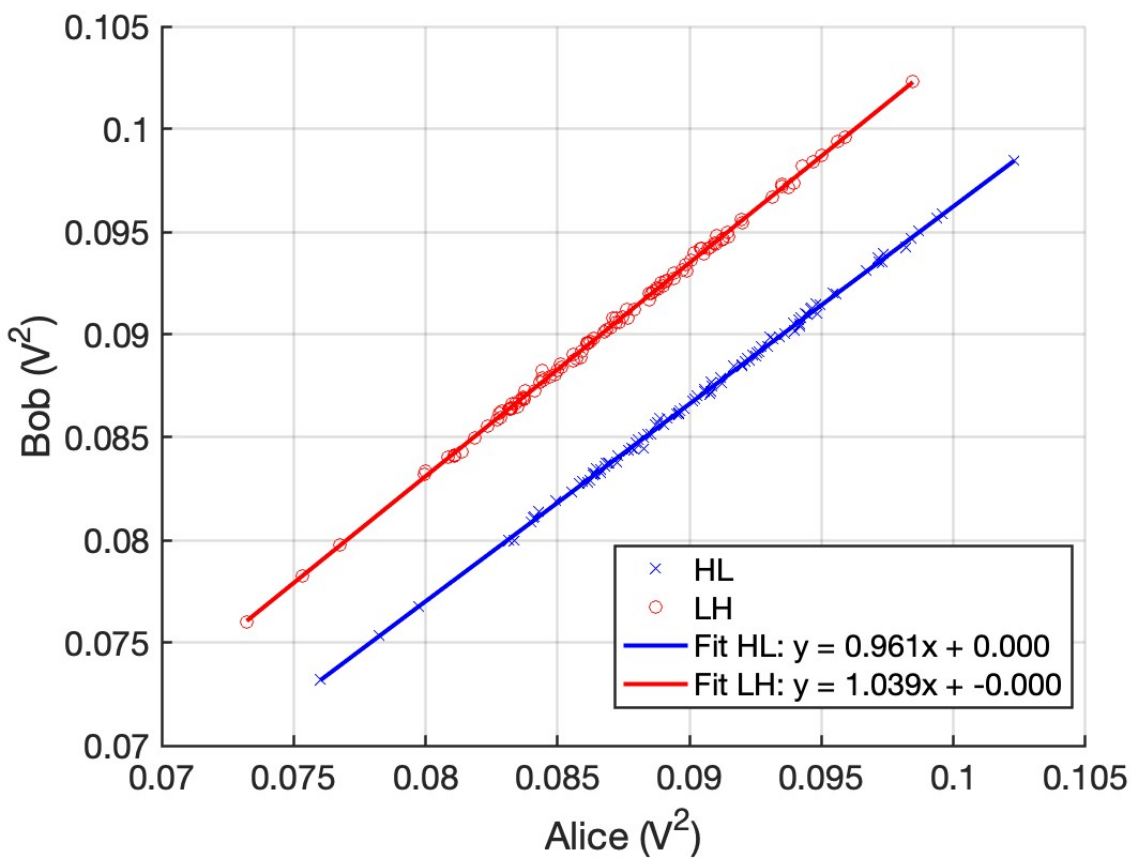

Figure 5. Illustration with a 200 Ω cable resistance that swapping the x and y axes for the LH data reproduces the curve for the HL case, thus reducing the number of independent measurements by half.

### *2.3 Defending the KLJN system against the new attack by the Vadai-Gingl-Mingesz method*

We applied the VGM method [52] for the temperature ratio of the L and H sides, see Figure 6. It can be clearly seen in the figure that the VGM method nullifies this attack. The observable minuscule difference of the slopes of the fitted lines does not carry information leak because it is random as it is due to the small data set available for Eve during the BEP duration.

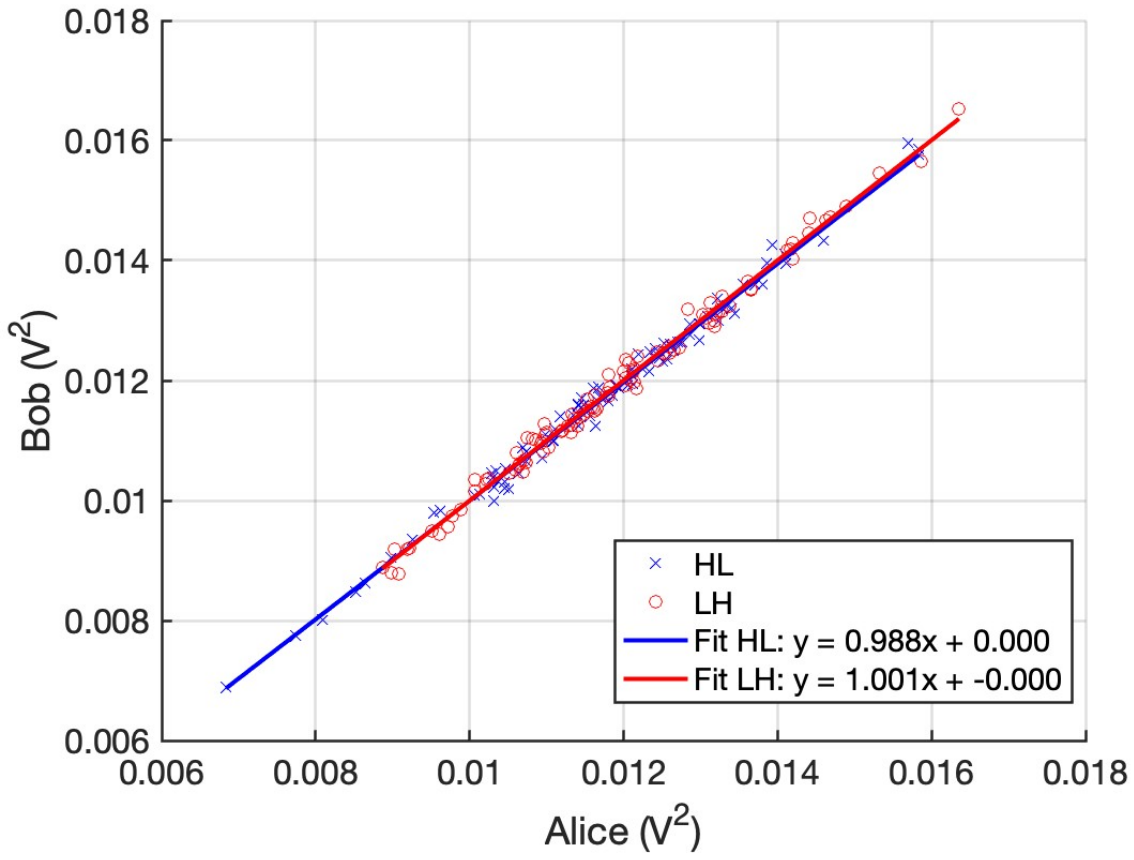

Figure 6. Complete defense against of the new attack at 200 Ω wire resistance by the VGM type noise temperature modification, compare to Figures 3 and 5. The observable minuscule difference of the slopes of the fitted lines does not carry information leak because it is random as it is due to the small data set available for Eve during the BEP duration.

The same conclusion can be drawn in the case of the large (800 Ω) wire resistance case, see Figure 7.

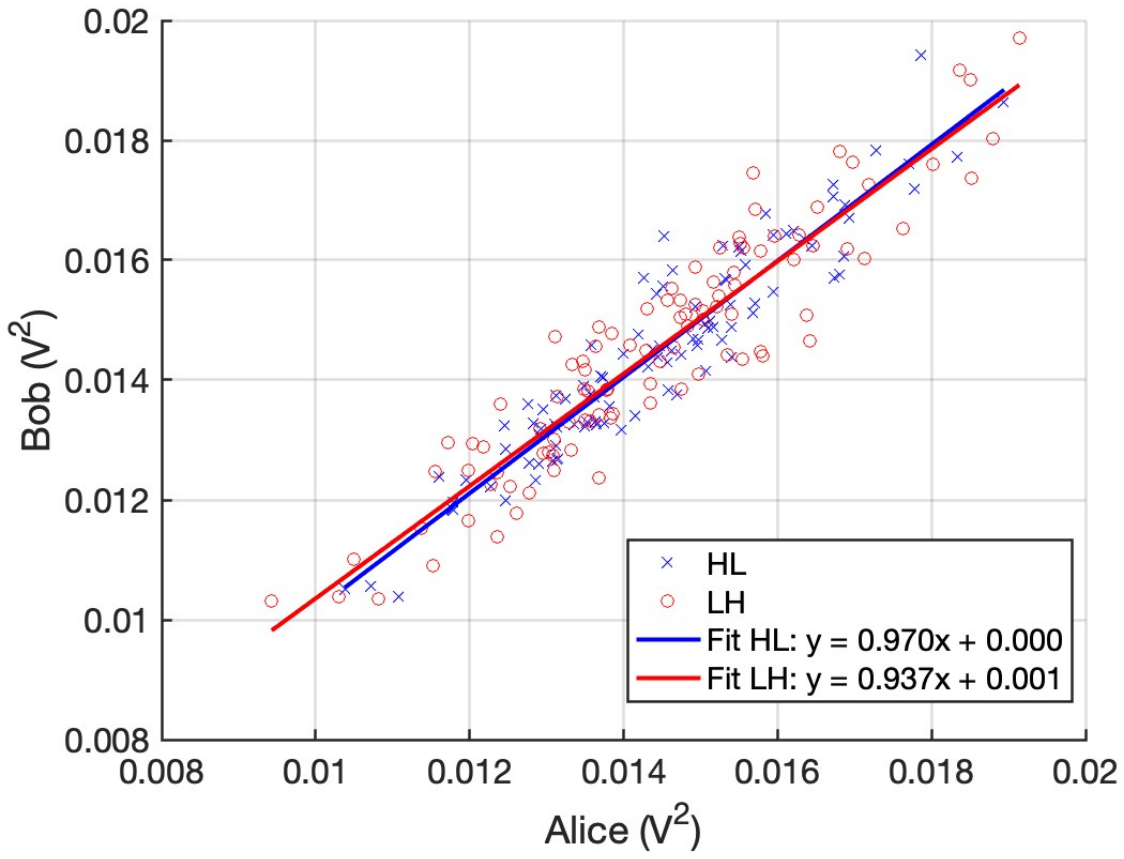

Figure 7. Illustration of the success of the VGM defense at a very large, 800 Ω, wire resistance by their noise temperature modification, compare to Figure 4. The conclusion is the same as at Figure 6.

However, one should note that, unfortunately, such a change of temperatures increases the information leak in the cable capacitance attack [91].

## 3. Conclusion

The most efficient wire-resistance-based attack identified so far targets the KLJN key exchange protocol by exploiting the correlated statistical fluctuations observed at the two ends of the communication line. These fluctuations are represented on a two-dimensional plot, where their values serve as the x and y coordinates. This visualization effectively functions as a binary classifier: the resulting curves, which are nearly straight lines, exhibit distinct slopes for the LH and HL configurations. When the wire resistance is nonzero, the attack allows Eve to extract the secure bits with an almost 100% success rate.

The only known defense against this attack is adjusting the noise temperature using the correctly derived VGM factor, either increasing it at the low-resistance end or decreasing it at the high-resistance end [52]. However, this countermeasure has the drawback of improving Eve's signal-to-noise ratio in the cable-capacitance attack , thus introducing a design trade-off. Moreover, it remains an open question whether the cable-capacitance attack [91] could be further strengthened by exploiting similar correlated statistical fluctuations at both ends of the channel.

Finally, it is still uncertain whether other effective defense mechanisms can be developed against this newly identified attack.

**Acknowledgements**

LBK is grateful for a valuable discussion to Kevin Nowka.